\documentclass[10pt,conference,letterpaper]{RWWTemplate}
\IEEEoverridecommandlockouts

\usepackage{times,amsmath,amssymb,graphics,graphicx}

\title{Novel Baseband Equivalent Models of Quadrature Modulated All-Digital Transmitters}

\author{%
Omer Tanovic{\small $~^{1,2}$},  Rui Ma{\small $~^{1}$}, 
and Koon Hoo Teo{\small $~^{1}$}%
\vspace{12pt}\\
$~^{1}$Mitsubishi Electric Research Laboratories, Cambridge, MA, 02139, USA, 
rma@merl.com\\
$~^{2}$Laboratory for Information and Decision Systems (LIDS), Massachusetts Institute of \\
Technology (MIT),  Cambridge, MA, 02139, USA
\thanks{This work was done while Omer Tanovic was an intern at Mitsubishi
Electric Research Laboratories (MERL).}}

\begin{document}
\maketitle

%
\begin{abstract}

In this paper an exact baseband equivalent model of a quadrature modulated all-digital 
transmitter is derived. No restrictions on the number of levels of a digital switched-mode 
power amplifier (SMPA) driving input, nor the pulse encoding scheme employed, are made. 
This implies a high level of generality of the proposed model. We show that all-digital 
transmitter (ADT) can be represented as a series connection of the pulse encoder, 
discrete-time Volterra series model of fixed degree and memory depth, and a linear 
time-varying system with special properties. This result suggests a new analytically 
motivated structure of a digital predistortion (DPD) of SMPA nonlinearities in ADT. 
Numerical simulations in MATLAB are used to verify proposed baseband equivalent model. 

\end{abstract}

\section{Introduction}

Behavioral modeling of nonlinearities in power amplifiers (PA), and their compensation, 
has been an active area of research for more than two decades (see \cite{Kim2001}-
\cite{Rawat2014} and references therein). Most of the work has focused on traditional 
transmitters employing linear PAs. All-digital transmitters (ADT), which use switched-mode 
power amplifiers (SMPA) instead of linear PAs, have recently gained significant importance 
due to their promising potential of providing high power efficiency (see \cite{Raab2003}-
\cite{Raab2010} and references therein). Surprisingly, not much research has been done 
on behavioral modelling of all-digital transmitters. Neural networks based behavioral ADT 
models were presented in \cite{Routsalainen2014}. Main deficiency of such black box 
modeling is the lack of insight into structure and internal dynamics of the nonlinear system 
being approximated. 
An exact baseband equivalent model for a quadrature modulated ADT was given in 
\cite{Routsalainen2015}, but only for 2 level pulse encoding scheme (taking precisely 
$\pm1$ values), and without an obvious extension to multi-level case. An analog 
baseband equivalent model was recently given in \cite{Ding2016}. Effects of pulse encoder 
are completely ignored in the derivation of equivalent model, thus neglecting large amount 
of quantization noise in the PA input, which spectrally spreads into signal band once 
processed by the passband nonlinearity. 

Recently, in \cite{Tanovic2016a}-\cite{Tanovic2016b}, authors showed that passband 
nonlinearities can be represented by an equivalent baseband model consisting of a series 
connection of a short memory nonlinear system followed by a special linear time-invariant 
system (LTI). They consider quadrature modulated transmitters in traditional setup, 
employing analog upconversion to carrier frequency and linear PAs. Assumption which 
enables such a compact system description is the zero-order hold digital-to-analog conversion 
of a PA driving signal. By recognizing similar form of the SMPA driving input, we extend that 
description to all-digital transmitters. 
Proposed ADT structure does not depend on the choice of pulse encoder (e.g. pulse-width 
modulation or delta-sigma modulation), 
nor does it depend on the number of its output levels. It exactly represents ADT when the 
passband nonlinearity can be represented by continuous-time Volterra series of fixed memory. 
Furthermore, it suggests a novel, non-obvious, analytically motivated DPD structure, which 
reduces computational and hardware complexity of a standard memory polynomial or pruned 
Volterra-series based DPD. 

\section{Problem Formulation}

Detailed block diagram of a quadrature modulated ADT is shown in the figure below. 
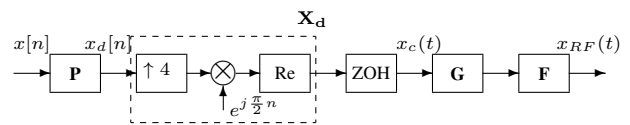
\begin{figure}[h]
\setlength{\unitlength}{0.082cm}
\scriptsize
\begin{center}\begin{picture}(98,22)(0,45)
\put(0,54){\vector(1,0){6}}
\put(6,51){\framebox(8,6){\bf P}}
\put(14,54){\vector(1,0){6}}
\put(20,51){\framebox(8,6){$ $}}
\put(21,53.8){$\uparrow4$}
\put(28,54){\vector(1,0){4}}
\put(34,54){\circle{4}}
\put(33,53){\line(1,1){2}}
\put(33,55){\line(1,-1){2}}
\put(32,53){\large$\times$}
\put(34,48){\vector(0,1){4}}
\put(36,54){\vector(1,0){4}}
\put(40,51){\framebox(8,6){Re}}
\put(48,54){\vector(1,0){6}}
\put(54,51){\framebox(8,6){ZOH}}
\put(62,54){\vector(1,0){6}}
\put(68,51){\framebox(8,6){\bf G}}
\put(76,54){\vector(1,0){6}}
\put(82,51){\framebox(8,6){\bf F}}
\put(90,54){\vector(1,0){6}}
\put(0,58){$x[n]$}
\put(11.5,58){$x_d[n]$}
\put(35,47){$e^{j\frac\pi2n}$}
\put(19,46){\dashbox(30,14){$ $}}
\put(46,62){$\bf X_d$}
\put(62,58){$x_c(t)$}
\put(88,58){$x_{RF}(t)$}
\end{picture}\end{center}
\caption{Detailed block diagram of a quadrature modulated ADT.}
\label{RWW_ADT_detailed}
\end{figure}
High resolution input sequence $x=x[n]$, whose in-phase and quadrature components 
are denoted $i=i[n]$ and $q=q[n]$ respectively, is first processed with a pulse encoder, 
denoted $\bf P$, so to assume amplitude levels suitable for driving a low resolution SMPA. 
Here we assume that $\bf P$ acts on $i$ and $q$ separately, and produces digital output 
signal $x_d=x_d[n]=i_d[n]+jq_d[n]$, where $i_d = {\bf P}i$, and $q_d = {\bf P}q$. 
Signal $x_d$ is first upconverted to a digital carrier frequency, before being converted to 
continuous-time domain by a zero-order hold (ZOH) digital-to-analog converter (DAC). 
Upconversion to digital frequency is usually done at the 
sampling rate of 4 times the carrier frequency. Under this assumption, mixing in digital 
domain becomes equivalent to interleaving the in-phase and quadrature components, 
together with their negative counterparts. We denote the digital upconversion system 
as $\bf X_d$ (Fig. \ref{RWW_ADT_detailed}). Continuous-time (CT) pulsed signal 
$x_c=x_c(t)$ is fed into SMPA, which is denoted  by $\bf G$ in Fig. 
\ref{RWW_ADT_detailed}. We are particularly interested in the case when $\bf G$ can 
be described by the general CT Volterra series model \cite{Schetzen2006}, with memory 
not larger than the sampling period of signal $x_d$. 

Output of the nonliner system $\bf G$ is passed through an analog bandpass filter 
$\bf F$, in order to produce the RF signal $x_{RF}=x_{RF}(t)$ to be radiated by 
antenna. Here we assume $\bf F$ to be an ideal bandpass filter with bandwidth 
equal to the sampling rate of the input signal $x$, and centered at the carrier 
frequency $f_c$. This is to avoid modeling the whole range of the output signal's 
spectrum which would make the linearization bandwidth very large, and would put 
a significant burden on system design. 

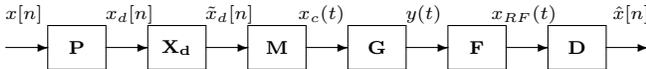
\begin{figure}[h]
\setlength{\unitlength}{0.095cm}
\scriptsize
\begin{center}\begin{picture}(109,9)(0,52)
\put(0,54){\vector(1,0){6}}
\put(6,51){\framebox(8,6){$ \bf P$}}
\put(14,54){\vector(1,0){6}}
\put(20,51){\framebox(8,6){$ \bf X_d$}}
\put(28,54){\vector(1,0){6}}
\put(34,51){\framebox(8,6){$\bf M$}}
\put(42,54){\vector(1,0){6}}
\put(48,51){\framebox(8,6){$ \bf G$}}
\put(56,54){\vector(1,0){6}}
\put(62,51){\framebox(8,6){$ \bf F$}}
\put(70,54){\vector(1,0){6}}
\put(76,51){\framebox(8,6){$ \bf D$}}
\put(84,54){\vector(1,0){6}}
\put(0,58){$x[n]$}
\put(14,58){$x_d[n]$}
\put(28,58){$\tilde x_d[n]$}
\put(41,58){$x_c(t)$}
\put(56,58){$y(t)$}
\put(68,58){$x_{RF}(t)$}
\put(85,58){$\hat x[n]$}
\end{picture}\end{center}
\caption{Series connection of ADT and a demodulator system.}
\label{RWW_System_S}
\end{figure}

ADT, as a system, maps baseband DT signal $x$ into passband CT signal $x_{RF}$. 
If baseband equivalent model is used for DPD design, it is more beneficial to find the 
map between $x$ and the downconverted and sampled version of $x_{RF}$, which 
we denote as $\hat x$. Therefore, instead of considering the ADT system given in 
Fig. \ref{RWW_ADT_detailed}, it is natural to consider a system depicted in Fig. 
\ref{RWW_System_S}, and denoted $\bf S$. Here $\bf D$ denotes an ideal 
demodulator, which runs at the sampling rate of baseband input signal $x[n]$, and 
$\bf M$ is used to denote the ZOH converter.

\noindent 
The SMPA driving signal $x_c$, i.e. input into $\bf G$, is a piecewise constant signal, 
which assumes only a finite number of amplitude levels corresponding to the choice of 
the pulse encoder $\bf P$. This is a consequence of the zero-order hold operation of 
digital-to-analog converter $\bf M$. It is now easy to see that series interconnection 
$\bf DFGM$, in Fig. \ref{RWW_System_S}, has very similar structure to the analogous 
one in \cite{Tanovic2016a}. The only difference is that sampling rates of the input and 
output signals are not equal. Signal $\hat x$ runs at a sampling rate of the baseband 
signal $x$, while $x_d$ runs at a higher sampling rate. Let $K$ be the ratio of the 
sampling rates of $x_d$ and $x$. It was shown in \cite{Tanovic2016a}, 
that the equivalent of the composition system $\bf DFGM$ can be represented as a series 
interconnection $\bf LV$ of a Volterra type DT nonlinear system $\bf V$, and an LTI system 
$\bf L$. For the system shown in Fig. \ref{RWW_System_S}, the equivalent $\bf L$ should 
be a time-varying system, due to the above mentioned mismatch between input and output 
signal rates. In the following section we decompose this equivalent linear subsystem 
into a series interconnection of a multi-input single-output LTI subsystem $\bf L$ and a 
downsampler. 
\section{Main Result}

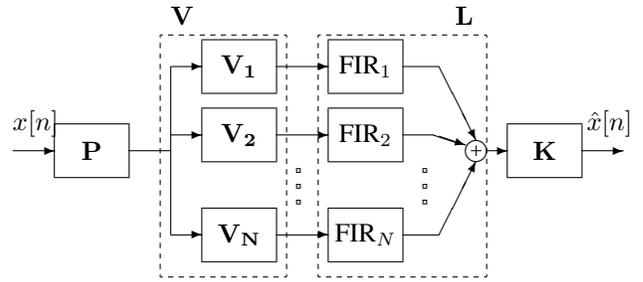
\begin{figure}[h]
\setlength{\unitlength}{0.070cm}
\begin{center}\begin{picture}(117,50)(0,30)
\put(30,70){\vector(1,0){6}}
\put(36,65){\framebox(14,10){$ \bf V_1$}}
\put(36,52){\framebox(14,10){$ \bf V_2$}}
\put(50,70){\vector(1,0){10}}
\put(60,65){\framebox(14,10){$\mbox{FIR}_1$}}
\put(60,52){\framebox(14,10){$\mbox{FIR}_2$}}
\put(50,57){\vector(1,0){10}}
\put(30,57){\vector(1,0){6}}
\put(22,54){\line(1,0){8}}
\put(8,49){\framebox(14,10){$ \bf P$}}
\put(0,54){\vector(1,0){8}}
\put(74,57){\line(1,0){6}}
\put(88,54){\circle{4}}
\put(88,55){\line(0,-1){2}}
\put(87,54){\line(1,0){2}}
\put(74,70){\line(1,0){7}}
\put(81,70){\vector(1,-2){7}}
\put(81,38){\vector(1,2){7}}
\put(60,33){\framebox(14,10){$\mbox{FIR}_N$}}
\put(74,38){\line(1,0){7}}
\put(80,57){\vector(3,-1){6}}
\put(78,50){\framebox(0.5,0.5){$ $}}
\put(78,47){\framebox(0.5,0.5){$ $}}
\put(78,44){\framebox(0.5,0.5){$ $}}
\put(54,50){\framebox(0.5,0.5){$ $}}
\put(54,47){\framebox(0.5,0.5){$ $}}
\put(54,44){\framebox(0.5,0.5){$ $}}
\put(50,38){\vector(1,0){10}}
\put(30,54){\line(0,1){16}}
\put(30,54){\line(0,-1){16}}
\put(30,38){\vector(1,0){6}}
\put(36,33){\framebox(14,10){$ \bf V_N$}}
\put(108,54){\vector(1,0){8}}
\put(109,58){$\hat x[n]$}
\put(0,58){$x[n]$}
\put(94,49){\framebox(14,10){$ \bf K$}}
\put(90,54){\vector(1,0){4}}
\put(28,30){\dashbox(24,46){$ $}}
\put(58,30){\dashbox(32,46){$ $}}
\put(30,78){$\bf V$}
\put(84,78){$\bf L$}
\end{picture}\end{center}
\caption{Baseband equivalent model of $\bf S$.}
\label{RWW_ADT_Equivalent_Block_Diag}
\end{figure}

Under assumptions introduced in the previous section, we give a general description of 
an ADT system. 
\vskip3mm
\noindent {\bf \sl Complete characterization of system $\bf S$ in Fig.\ref{RWW_System_S}}:\\
Let $\bf V$ be a single-input multi-output system mapping a complex signal $x[n]=i[n]+jq[n]$ 
into a vector valued real signal $v[n]=\begin{bmatrix} v_1[n] \dots v_N[n]\end{bmatrix}^T$, 
components of which are all Volterra monomials composed of $i[n]$ and $q[n]$, up to degree 
$M$ and depths $m_i$ and $m_q$. Let $\bf K$ be a downsampling by factor of $K$ subsystem, 
where $K$ was defined in the previous section. System $\bf S$, as given in Fig. \ref{RWW_System_S}, 
can be decomposed into a series connection of systems $\bf P$, $\bf V$, $\bf L$ and $\bf K$, 
as given in Fig. \ref{RWW_ADT_Equivalent_Block_Diag}, where $\bf L$ is a multi-input 
single-output LTI system with long memory, and good approximation by finite impulse response 
(FIR) filters (Proof is omitted due to space limitations.).  

Here we emphasize some consequences of this result. Block diagram representation of 
$\bf S$ in Fig. \ref{RWW_ADT_Equivalent_Block_Diag}, stresses parallel structure in the 
$\bf LV$ part of the system. Each $\bf V_k$ in Fig. 
\ref{RWW_ADT_Equivalent_Block_Diag} describes one Volterra monomial, e.g. 
$(V_kx)[n]=i[n]i[n-1]q[n]^2$. Therefore, subsystems $\bf V_k$ are relatively easy to 
implement since they only involve delays and multiplication operations. Attention should 
be payed to possible delays that can appear in subsystems $\bf V_k$. It can be shown 
that if continuous-time Volterra kernel has maximal memory $\tau$ (over all variables), 
then its corresponding discrete-time analog (group of monomials $\bf V_k$), has memory 
$\left\lceil\frac\tau T\right\rceil$, where $\lceil\cdot\rceil$ is a ceil function. Frequency 
responses of components of $\bf L$, depend heavily on the passband nonlinearity, but 
they have a common feature: they are smooth over the spectral band of interest. This 
makes them well approximable in spectral domain, by e.g. polynomials. Moreover, finding 
coefficients of those polynomials can be done by simple least squares optimization.  

It is important to emphasize that due to bandpass filtering of the PA output, long memory 
dynamic behavior is now present, which makes band-limited models fundamentally different 
from the conventional baseband models. This implies that traditional modeling methods, 
such as memory polynomials or dynamic deviation reduction-based Volterra series modeling, 
might be too general to define this new structure, and also not well suited for practical 
implementations due to long memory requirements (of the nonlinear part) which would 
require exponentially large number of coefficients. Therefore, the structure proposed in 
this paper does not suffer from issues of long memory of nonlinear subsystem, but rather 
has memory depth equivalent to that of the passband nonlinearity. Long memory 
requirements are forwarded to the linear subsystems $\bf L_k$, taking complexity burden 
off the nonlinear part.    

Now we discuss potential impact, of the above result, on digital predistortion design for 
all-digital transmitters. 
DPD design can be seen as a process of finding the inverse of $\bf S$ (when it exists). 
In most applications, with an appropriate scaling and time delay, the system $\bf S$ to 
be inverted can be viewed as a small perturbation of identity, i.e. $\bf S=I+\Delta$. 
Using Neumann's serries, assuming $\bf \Delta$ is small in an appropriate sense, the 
inverse of $\bf S$ can be well approximated by $\bf S^{-1}\approx I-\Delta=2I-S$. 
Hence the result of this paper suggests a specific structure of the compensator 
$\bf C\approx 2I-S$. As pointed out in \cite{Tanovic2016a}, this implies that plain 
Volterra monomials structure is, in general, not enough to model $\bf C$, as it lacks 
the capacity to implement the long memory effects caused by bandpass filtering. Instead, 
$\bf C$ should be sought in the form ${\bf C=I-KL_0}X{\bf VP}$, where $\bf P$ is the 
pulse encoder,$\bf V$ is the system generating all Volterra series monomials of a limited 
depth and limited degree, $\bf L_0$ is a fixed LTI system with a long time constant, $X$ 
is a matrix of coefficients to be optimized to fit the data available, and $\bf K$ is a 
downsampler. Elements of the coefficient matrix $X$ can be easily obtained by applying 
simple least squares optimization on the set of input output pairs of $\bf S$.  

\section{Simulation Results}

Validation of the above model of system $\bf S$ is done for two cases of passband 
nonlinearity $\bf G$. In the first case we assume a simple third order analog nonlinearity, 
consisting of a linear term and one CT Volterra monomial,  as given below
\begin{equation} \label{NL1}
(Gx)(t) = x(t) - \delta_1 x(t-\tau_1)x(t-\tau_2)x(t-\tau_3) .
\end{equation}
Time delays in the above expression are taken as $\tau_1=1.2T$, $\tau_2=2.3T$, 
$\tau_3=0.4T$, where $T$ is a sampling time of a digitaly upconverted DT signal 
$\tilde x_d$ (see Fig. \ref{RWW_System_S}). Parameter $\delta_1$ is varied in 
the interval [0.001,0.2]. This parameter roughly dictates the amount of passband 
distortion or equivalently the level of linearity of the transmitter circuit. 
In the second case we assume 
\begin{equation} \label{NL2}
(Gx)(t) = x(t) - \delta_2 \int_0^t h(\tau_1,\tau_2)x(t-\tau_1)x(t-\tau_2)d\tau_1d\tau_2,
\end{equation}
where kernel $h=h(t_1,t_2)$ is of memory less than $4T$, and is separable, i.e. 
$h(t_1,t_2)=h_1(t_1)\cdot h_2(t_2)$, with $h_1(t_1)=\exp(-0.95t_1)$ and 
$h_2(t_2)=\exp(-0.91t_2)\cos(\pi/5t_2)$. In this case, parameter $\delta_2$ takes 
values from [0.001,0.015]. In both cases, delta-sigma modulator (DSM) of first order and 
5 output levels, was used as pulse encoder. System $\bf S$ is modeled as given in Fig.  \ref{RWW_ADT_Equivalent_Block_Diag}. In both cases, systems $\bf V_k$ are Volterra 
monomials of maximal degree 3, and digital memory of up to 4 samples (due to maximal 
memory of passband nonlinearity equal to $4T$). Coefficients of the FIR filters are 
obtained by fitting this model on a training data of 40960 input-output pairs.

\begin{figure}
\center
\includegraphics[scale=.81]{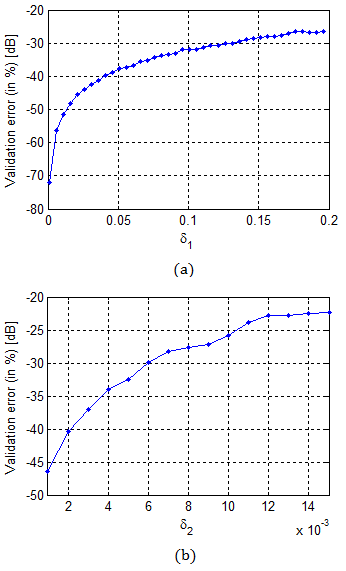}
\caption{Model Validation Error: (a) Example 1; (b) Example 2.}
\label{Model_Validation}
\end{figure}

Model validation results are depicted in Fig. \ref{Model_Validation}. Very good modeling 
error (below 0.1\% or -20dB) is achieved in both cases, though an upward trend in error 
(as $\delta_{1/2}$ is increased) is noticable from Fig. \ref{Model_Validation}. This is due 
to the error in approximating FIR filters, which gets enhanced as level of distortion 
increases.  


\section{Conclusion}
In this paper, we propose a novel structure of the baseband equivalent model of a 
quadrature modulated ADT, under assumption that the passband nonlinearity is of 
Volterra series type of fixed memory. It was shown that ADT can be represented as 
a series connection of the pulse encoder, discrete-time Volterra series model of fixed 
degree and memory depth, and a linear time-varying system with special properties. 
Model does not assume any particular pulse encoding method, nor does it put constraints 
on the number of its output levels. Results suggest a novel, analytically motivated DPD 
structure, which can potentially reduce computational and hardware complexity compared 
to traditional DPD design techniques. 




\end{document}